\documentclass[sigconf,anonymous=false]{acmart}

\AtBeginDocument{%
  \providecommand\BibTeX{{%
    \normalfont B\kern-0.5em{\scshape i\kern-0.25em b}\kern-0.8em\TeX}}}

\acmPrice{15.00}
\acmISBN{978-1-4503-XXXX-X/18/06}

\copyrightyear{2024}
\acmYear{2024}
\setcopyright{acmlicensed}\acmConference[Websci '24]{ACM Web Science Conference}{May 21--24, 2024}{Stuttgart, Germany}
\acmBooktitle{ACM Web Science Conference (Websci '24), May 21--24, 2024, Stuttgart, Germany}
\acmDOI{10.1145/3614419.3644006}
\acmISBN{979-8-4007-0334-8/24/05}

\setcopyright{none}
\renewcommand\footnotetextcopyrightpermission[1]{}
\makeatletter
\renewcommand\@formatdoi[1]{\ignorespaces}
\makeatother
\begin{document}

\title{Dynamics in Search Engine Query Suggestions for European Politicians}


\author{Franziska Pradel}
\email{franziska.pradel@tum.de}
\orcid{1234-5678-9012}
\affiliation{%
  \institution{Technical University of Munich}
  \city{Munich}  
  \country{Germany}
}

\author{Fabian Haak}
\email{fabian.haak@th-koeln.de}
\orcid{1234-5678-9012}
\affiliation{
  \institution{TH K\"oln (University of Applied Sciences)}
  \city{Cologne}
  \country{Germany}
}

\author{Sven-Oliver Proksch}
\affiliation{%
  \institution{University of Cologne}
  \city{Cologne}  
  \country{Germany}
}

\author{Philipp Schaer}
\affiliation{
  \institution{TH K\"oln (University of Applied Sciences)}
  \city{Cologne}
  \country{Germany}
}

\renewcommand{\shortauthors}{Pradel et al.}

\begin{abstract}
Search engines are commonly used for online political information seeking. Yet, it remains unclear how search query suggestions for political searches that reflect the latent interest of internet users vary across countries and over time. We provide a systematic analysis of Google search engine query suggestions for European and national politicians. Using an original dataset of search query suggestions for European politicians  collected in ten countries, we find that query suggestions are less stable over time in politicians' countries of origin, when the politicians hold a supranational role, and for female politicians.  Moreover, query suggestions for political leaders and male politicians are more similar across countries. We conclude by discussing possible future directions for studying information search about European politicians in online search.
\end{abstract}



\received{20 November 2023}

\maketitle

\section{Introduction}
The past decade has witnessed an increase in the politicization of the European Union in domestic politics as well as a rise in digitalization and online political information seeking. Prior studies have addressed this emerging digital European public sphere, a space that is needed in liberal democracies and which promotes a public debate between politicians and citizens in the EU~\cite{rivas-de-roca_understanding_2021}. 
However little is known about the cross-national aspects of online political information search related to EU issues and European politicians. Our study is the first to systematically investigate search query suggestions as latent indicators for information search related to European politicians and issues. Analyzing how the level of latent interest in politicians and political topics varies over time and across countries during the 2019 European Parliament election period can help understand drivers of political information search and politicization in Europe.

A common interest in political issues and European politicians should facilitate the establishment of a joint European public sphere. Studies have consistently stressed that there is a political communication gap in the EU \cite{marquart_knowing_2019}, the filling of which would be essential for the exchange and debate with EU citizens \cite{rivas-de-roca_understanding_2021}. While prior research focuses on mass-mediated communication, specifically on newspaper coverage \cite{mancini_countries_2015, hutter_politicizing_2019}, several scholars point to the relevance of examining digital platforms as an arena for EU politicization \cite{rivas-de-roca_understanding_2021, risse_european_2014, risse_introduction_2014}.

With regard to the political use of search engines, studies find that search query data can be used to predict issue salience \cite{scharkow_measuring_2011, mellon_internet_2014, swearingen_google_2014}, attitudes \cite{preis_quantifying_2012, pelc_googling_2013} and behavior \cite{arendt_googling_2019, stephens-davidowitz_cost_2014, ginsberg_detecting_2009, prado-roman_google_2021}. 
However, data with real-world user queries, query suggestions, and actual query reformulations are rare, mainly because of privacy concerns.
Query logs allow for the identification of users, despite pseudonymization techniques and thus are no longer collected, published, or analysed~\cite{Barbaro_aol}.

In this study, we compare political search query suggestions across borders to estimate latent interest in European politics. Query suggestions (also called search predictions~\cite{sullivan_2018}, query auto completion~\cite{cai_survey_2016}, or query suggestion predictions) are the list of search queries provided to a search engine user when entering a query into a search bar. They are based on the queries searched for in the same region and language containing the query the user has input so far \citep{google_search_help}. Query suggestions imply what users search for and thus serve as indicators of interest in the given query topic, allowing for a comparative analysis of latent interests in multiple countries. 
Google Trends data does not serve this purpose. Since Trends data can only be retrieved as relative information (relative to the maximum value of all retrieved data), multiple searches cannot be merged. Thus, Google Trends cannot be used to compare a larger number of countries and politicians~\cite{scharkow_measuring_2011}. 
We introduce a novel dataset on Google query suggestions for the names of European politicians and political issues as base queries. 
The dataset is assembled by collecting query suggestions from Belgium, the Czech Republic, France, Germany, Italy, Netherlands, Norway, Portugal, Spain, and the United Kingdom. Focusing on the query suggestions in ten European countries allows us to examine and compare latent political interest approximated with the stability of Google query suggestions for politicians over time and similarity across European countries. 
We test several determinants for explaining the stability of query suggestions for politicians within countries and the differences across countries. 

Our findings suggest that the gender,  role, government status, and country of origin of politicians significantly impact the query suggestions shown to online users. Query suggestions for the names of politicians are less stable over time in the politicians' country of origin, when the politicians hold a supranational political role, and when they are female. These findings suggest an increased latent interest in these politicians and users searching for more diverse information online. 
Thus, the findings of this interdisciplinary study provide novel insights into users' information-seeking behavior by suggesting that the online interest and engagement continues to be rooted in national contexts. 

\section{Related Research and Theoretical Background}

\subsection{Search Query Suggestions}\label{sec:QS_theory}
Search engines provide the most popular approach to finding information online, although conversational information-seeking tools are getting more popular~\cite{INR_081}. 
Search engines are perceived as trustworthy sources of information on many topics, including political information~\cite{ray_2020,edelman_2020}. \citet{epstein_search_2015} and \citet{Houle_2015} have shown that search engines' results have the potential to influence users' political opinions and even voting behavior. 
Query suggestions (also called query suggestions, search suggestions), the list of predictive completions provided by search engines during the input of a search query, play a crucial role in what people search for~\cite{Niu_2014}.
Query suggestions are primarily interacted with to save time during the input of a query or if the information need of users is less established~\cite{Niu_2014}.
In the case of political information, the latter can impact the information formation of users by guiding information exposure.

Search engine providers keep the exact mechanisms behind query suggestion features obscure. However, the main principles behind them are known. As stated by the most popular search engine company Google, predictions are based primarily on users' previous searches on the same topic in a similar location ~\cite{google_search_help,sullivan_how_2018}.
In our dataset, information on the location and language of the search are provided to the search engine along with the search term via HTTP request.

Google states that potentially harmful, defaming, or unwanted predictions are filtered based on auto-complete policies \citep{google_2021}. Nevertheless, we know neither how exactly query suggestions are filtered nor whether this happens automatically or by removal request (c.f., \cite{google_removal}).
Due to the dependency on trends in searches both temporally and spatially, query suggestions can be seen as indicators of what people search for. 
Wang and colleagues have shown that query suggestions can be manipulated by deliberately submitting searches \cite{wang_2018}. Therefore, we can assume that query suggestions are an approximate representation of peoples' information needs.

\subsection{Political Information in the Digital Society}

The internet, search engines, social media, and other online services have transformed how people find political information. The ability of citizens in democracies to access political information freely at any time online is unprecedented \cite{bimber_information_2003, swanson_political_2003, jungherr_retooling_2020, stockemer_internet_2018}. 
Search engines play an important role for people who want to inform themselves about political issues and politicians~\cite{trevisan_google_2018, pradel_biased_2020}. They often serve as the primary and initial source of information~\cite{dutton_search_2017}. 
There is an interconnection between both different online platforms as well as between them and traditional media~\cite{chadwick_hybrid_2017}. For instance, politicians and political events receive public attention both on social media platforms and in traditional media like newspapers~\cite{jungherr_retooling_2020}, and both can reciprocally influence visibility \cite{kruikemeier_understanding_2018}. Encountering an unknown politician or a political event, for example, on television or social media, can trigger individuals to go online and search for the event in a search engine like Google~\cite{trevisan_google_2018}. 

A search engine inherently serves as a gatekeeper, exerting significant influence over the political information that users encounter. Websites that appear at the top of search results attract more interactions~\cite{dutton_search_2017}. While there is no conclusive evidence yet as to whether search engine results affect voting decisions in Europe, some experimental evidence suggests that when politicians are ranked higher in search engine results, they are more likely to be favored by undecided voters, an effect that is described as the search engine manipulation effect~\cite{epstein_search_2015}.  

Google is Europe's most popular search engine, with a 92\% market share (November 2023) \cite{statcounter2023}.
Query suggestions are an essential feature of search engines that can direct a user's information search.
However, the query suggestions can be systematically biased based on the politicians' characteristics like gender and party~\cite{pradel_biased_2020, bonart_investigation_2019, HaakS21, Haak_2022}, just as other political information on online platforms such as Twitter/X~\cite{mertens_as_2019, spinde_what_2023} and Wikipedia can be systematically biased~\cite{pradel_biased_2020}.

\subsection{Interest in European Politicians Across Borders}\label{sec:Interest}

With the increasing policy-making role of the European Union and repeated common crises in Europe (e.g., the financial crisis, the refugee crisis, the climate crisis, and the Corona pandemic), more and more public attention is dedicated to events of pan-European consequences. At the same time, European integration has become politicized across Europe. New challenger parties on the left and on the right mobilize the electorate around this new transnational cleavage~\cite{hobolt_issue_2015, hooghe_cleavage_2018}. While the European elections had previously been described as ``second-order elections''~\cite{reif_nine_1980} due to low voter turnout and the success of extreme parties, a reversal in turnout in the 2019 elections and the nomination of lead candidates, also called \textit{Spitzenkandidaten}, during the European election campaign has installed a much stronger European element into European election campaigns~\cite{schmitt_it_2020,Kotanidis2023}. In the process of lead candidate nomination, European parties select a lead candidate that they would support for the office of President of the Commission to, among other things, energize the EU elections through a personalized campaign and public debates aiming at increasing the voter turnout \cite{Kotanidis2023}. However, the presentation of lead candidates differed between countries, and they received low media attention in European countries, for example, in British, French, and German newspapers during the 2014 campaign~\cite{schulze_spitzenkandidaten_2016}. 

People obtain their information on international politics through mass media, and the way the latter report consequently influences public opinion about other countries~\cite{balmas_leaders_2013}. The relationship between states can ultimately be influenced by an understanding of countries and people and even the perceived legitimacy of countries~\cite{balmas_leaders_2013}. Yet, investigating the news coverage of European politicians is rather complex, as Europe is characterized by having heterogeneous media systems, and citizens do not get their information from a single source~\cite{hallin_comparing_2004}. Furthermore, media consumption habits vary between European countries~\cite{castro_navigating_2021}.

Studies show that despite EU politics and politicians being relatively underrepresented in the European media, their visibility is higher around EU summits and in countries with higher satisfaction in democracy~\cite{peter_search_2004}. 
Research on the representation of foreign leaders shows that not only media coverage of national politics~\cite{balmas_two_2014, adam_personalization_2010} but also foreign politics~\cite{balmas_leaders_2013} is becoming increasingly personalized in Western democracies. 
This implies that the media increasingly focuses on foreign political leaders' personalities rather than on the nations as such~\cite{balmas_leaders_2013}. 

This raises the question of how users search for information about European politicians, especially political leaders. We analyze the stability and cross-national similarity of Google search query suggestions that we collected in  Belgium, Czech Republic, France, Germany, Italy, Netherlands, Norway, Portugal, Spain, and the United Kingdom over a two-year period which included the European election campaign in 2019.

\section{Expectations about Query Suggestions as Latent Interest Indicators}\label{sec:expectations}

\subsection{Stability of Query Suggestions}
In our study, the stability of search query suggestions, measured as the degree of change of suggestion over time, serves as an indicator of latent interest. We assume that higher interest coincides with more diverse search queries that fluctuate more over time. Arguments from research on media consumption support this assumption, suggesting that people who are highly (politically) interested in a subject seek diverse perspectives and information on the subject \cite{yuan_2011, dubois_2018}.  In Section~\ref{sec:results_stability}, we further demonstrate the validity of this assumption by showing that 
we indeed observe more fluctuations in query suggestions for critical scenarios where strong theoretical reasons would also anticipate more interest. We investigate stability for a range of meta-attributes (e.g., political ideology, gender, political role) for which we expect specific effects discussed in this section.

Previous research has stressed that citizens have more interest in national politicians and national elections \cite{reif_nine_1980, bright_europes_2016}. We therefore expect more diverse online information searches for these politicians and less stable query suggestions over time. 
\textit{Accordingly, we hypothesize that query suggestions for politicians are less stable when the search for politicians is carried out in their country of origin (H1).}  

Political ideology may further affect the stability of query suggestions. As more extreme parties that challenge the established party system seek and attract more attention in the electorate and media~\cite{boomgaarden2007explaining, walgrave2004making}, we may hypothesize that internet users are also more likely to search for specific events surrounding the political strategies of extreme parties.  \textit{Therefore, we hypothesize that centrist parties' query suggestions are more stable than those of extreme parties (H2).}
We include left-right ideology and a squared term from the Chapel Hill expert survey on party positions~\cite{bakker_chapel_2019} to capture policy extremity.

The interest in politicians may also vary with their political role. Party leaders of national parties receive more frequent media coverage than other politicians, particularly through the increasing focus on single personalities in politics and the media, and they are increasingly critical for voting decisions \cite{karvonen2010personalisation, takens2015party, rahat2007personalization}. 
Such increased media coverage could, in turn, lead to increased search interest and more diverse search queries over time. 
Comparatively, lead candidates with a supranational role receive less media attention in their country of origin~\cite{schulze_spitzenkandidaten_2016}.
However, due to their supranational role, we would expect more average media attention across European countries.
We expect that the stability is the lowest when the politicians have a political leader role: \textit{We hypothesize that the stability of query suggestions is lower for party leaders compared to members of the national cabinet~(H3) and of lead candidates compared to members of the national cabinet~(H4).}

We further explore the role of gender. Search engine query suggestions can show topically biased political information based on politicians' gender, which has been demonstrated on a national level~\cite{pradel_biased_2020, HaakS21, Haak_2022, bonart_investigation_2019}. 
Specifically, voters' information-seeking varies by politicians' gender with, for instance, more competence-related information searches when the politician is female~\cite{ditonto_gender_2014}. As this study experimentally revealed that individuals seek more information about female politicians, it might be the case that internet users also seek more information about female politicians in search engines. If female politicians are much more controversially debated in public, this would be even more true, as users' search interest is triggered by a constant flow of new, diverse information due to ongoing monitoring in media and the public. 
We would assume to find these gender bias effects in our study on an international level, leading to higher volatility in query suggestions for female politicians and more stable query suggestions for male politicians. 
Therefore, we test whether the stability of query suggestions varies by politicians' gender in the European arena:  
\textit{We expect that query suggestions for male politicians are more stable than those for female politicians~(H5).} 

 Another more systemic determinant for more diverse query suggestions is the general salience of the EU in the country where the search is performed. If the EU is a salient issue, there may be more interest in other European politicians within these countries, which impacts the query suggestions.
\textit{We hypothesize that the query suggestions are less stable the higher the EU salience~(H6).}  

Finally, we also assume that the government status of the politician plays a major role. The politics and politicians of the government party are more frequently covered in mass media~\cite{10.1111/j.1460-2466.2011.01540.x}. Moreover, citizens might seek more diverse information about politicians over time when they are in the government party. \textit{Accordingly, we expect that the query suggestions are less stable for politicians in the government party than those in the opposition party~(H7).}  

\subsection{Similarity Across Countries}

In addition to examining stability over time, we explore how much query suggestions vary cross-nationally.
We expect that mainstream centrist parties are more likely to have similar query suggestions than extreme parties across countries, as politicians from extreme parties should also have higher volatility within their country. \textit{Therefore, we hypothesize that the query suggestions for centrist politicians are more similar across countries than those belonging to extreme parties~(H8).}
We capture this relationship with two variables: left-right ideology and its squared term. 

Moreover, we expect that the similarity of the query suggestions varies with the politician's role. Query suggestions for European-wide lead candidates may be more similar compared to national party leaders and members of the cabinet because they are supranational actors, and there might be a common perception and interest in them.
\textit{We hypothesize that the similarity of query suggestions across countries is higher for lead candidates than party leaders and members of the national cabinet~(H9).}

As before, we expect the gender of the politician to play a significant role when explaining the cross-national similarity of query suggestions. We expect that the information searches are more similar across countries for male politicians because different gender roles and attitudes toward gender equality across countries \cite{alesina2013origins, paxton2020women} could affect the search behavior. \textit{Thus, we expect that the query suggestions are more similar across countries for male politicians than for female politicians~(H10).}
Finally, it might be the case that being in the government party also influences the similarity of query suggestions as these politicians have a considerable impact on policies and thus might receive more attention. Searches for members of government parties may therefore be more context-specific and should differ more across countries. \textit{Accordingly, we hypothesize that being in the government party decreases the cross-national similarity of query suggestions~(H11).}

\section{Comparative Analysis of European Search
Query Suggestions}\label{sec:comp_anal}

\subsection{Data Collection}
For the analyses, we collected query suggestions for selected politicians between the beginning of 2019 (January 18) and the end of 2020 (December 18), including the election period for the European Parliament in May 2019.\footnote{Twice per day, a specialized web crawler has collected query suggestions by HTTP request from Google's autocomplete API. The returned lists of query suggestions are stored in an SQL database along with the corresponding search terms.} 
The  dataset consists of more than 46.64 million query suggestions for the names of 793 politicians (including 14 lead candidates and nominees of the European election 2019, 261 politicians with a political leadership role, and 518 members of national cabinets of all 27 member states of the European Union). 
Lead candidates of the European parties are also often called \textit{Spitzenkandidaten} and refer to the European lead candidates of European parties for the European elections, for instance, Manfred Weber and Ska Keller, while nominees of the lead candidate position refer to politicians being nominated in the first round as lead candidates but were not selected (e.g., Petra De Sutter, Atanas Schmidt). 
The information was derived from lists on Wikipedia or official websites of political cabinets. In total, the dataset covers 226 female and 567 male politicians, with an average political ideology score of 5.425 on a scale ranging from 0 (extreme left) to 10 (extreme right), which has been obtained from the 2019 CHES expert survey on political parties~\cite{bakker_chapel_2019} (c.f., Section~\ref{sec:analytical_approach}). Table \ref{tab:politicians_gender_representation} gives an overview of the different types of our data.

\begin{table}[h]
\caption{Overview of Politicians and Gender Representation}
\begin{tabular}{p{4.7cm}p{1.39cm}p{1.34cm}} \toprule
Type of politician & Observations & Female (\%)\\ \midrule %
Lead candidates of European parties in 2019 EP election & 5 & 20.0\% \\
Lead candidate nominees of European parties in 2019 EP elections & 9 & 44.4\% \\
Party leaders from 27 countries & 261 & 23.4\% \\ 
Cabinet ministers from 27 countries & 518 & 30.9\% \\ \midrule 
Observations & 793 & 28.5\% \\ 
\bottomrule
\end{tabular}
\label{tab:politicians_gender_representation}
\end{table}

Via proxy servers, the search was performed in various European countries, i.e., Belgium, Czech Republic, France, Germany, Italy, Netherlands, Norway, Portugal, Spain, United Kingdom.
Using Google Translate, all query suggestions were translated into English, allowing us to compare the similarity of query suggestions for politicians across countries.\footnote{The dataset is available at Zenodo \url{https://doi.org/10.5281/zenodo.10693919}.}

Table \ref{suggestions_example_merkel} shows translated query suggestions for the German chancellor at that time, Angela Merkel, as an example. 
The query suggestions of the 6\textsuperscript{th} of June 2019 and of the 20\textsuperscript{th} of June 2019 differ more than the suggestions between the 18\textsuperscript{th} of January and the 1\textsuperscript{st} of February. The larger change coincides with a public health incident of Angela Merkel who was visibly trembling at a public event on the 18\textsuperscript{th} of January. The event was covered by national and international news \cite{der_spiegel_zittern_2019, die_welt_merkel_2019, le_figaro_angela_2019, el_mundo_angela_2019, bbc_germanys_2019}.
Table \ref{tab:example_merkel_countries} shows examples of  query suggestions in four countries for Angela Merkel on the 20\textsuperscript{th} of June 2019. 
Query suggestions concerning her health appear in the German query suggestions. While these query suggestions also appear in French search query suggestions, they are absent from Belgian (Dutch) and Portuguese suggestions. We can also see how suggestions reflect national topics in the non-German suggestions. For example, in Portugal, the search suggestion of ``madeira'' might refer to a deadly bus crash in April 2019 mainly involving German tourists, for which Angela Merkel issued her condolences~\cite{schuetze_merkel_2019}. 
While Table~\ref{suggestions_example_merkel} illustrates that salient events can quickly appear in query suggestions of a politician's country of origin, Table~\ref{tab:example_merkel_countries} shows that the suggestions of a politician can vary across countries.

\begin{table}
\caption{Exemplary query suggestions for Angela Merkel in Germany. Note: The authors have translated the query suggestions.}
{\begin{tabular}{p{6cm}p{1.75cm}} \toprule
 Query suggestions in Germany & Date \\ \midrule
young, curriculum vitae, husband, news, size, kassel, formerly, salary, profile & 2019-01-18 \\
young, salary, curriculum vitae, apartment, news, size, assets, profile, formerly   & 2019-02-01 \\
\multicolumn{1}{c}{...}                                                                               & \multicolumn{1}{c}{...} \\
children, age, harvard, mother, salary, husband, young, apartment, assets           & 2019-06-06 \\
trembles, national anthem, sick, children, age, salary, parkinson, trembling, today    & 2019-06-20 \\
\bottomrule
\end{tabular}}
\label{suggestions_example_merkel}
\end{table}

\begin{table}[t]
\caption{Exemplary query suggestions for Angela Merkel for different countries on June 20\textsuperscript{th},  2019 (translated by the authors).}
{\begin{tabular}{p{6cm}p{1.75cm}} \toprule
 Query suggestions  & Country \\ \midrule
trembles, national anthem, sick, children, age, salary, parkinson, trembling, today                                                  & Germany             \\
lyrics, gestapo lyrics, divorced, clothing, wikipedia, young, wiki, twitter, stops                                              & Belgium (Dutch) \\
young, age, sick, trembling, faint, twitter, video, biography, husband                                                           & France              \\
twitter, party, children, age, wiki, wikipedia, madeira, ulrich merkel, biography & Portugal \\

\bottomrule
\end{tabular}}
\label{tab:example_merkel_countries}
\end{table}

\subsection{Measuring Stability of Query Suggestions}
We use the Jaccard similarity coefficient to measure query suggestions' stability over time.
The coefficient varies between 0 and 1, with 0 indicating no overlap of the query suggestions and 1 indicating that they are identical.  

While we collect daily query suggestions for each politician, we aggregate the query suggestions to a two-week level for our analysis. We chose this period as it allows us to capture potential changes in query suggestions. Changes are hardly detectable if the period is shorter than two weeks.
Moreover, despite events taking time to be distributed, first by national and then foreign media, and then taking time until query suggestions reflect these events, we do not expect events to take more than two weeks to appear in query suggestions.
We utilize the Jaccard coefficient to reveal the stability of query suggestions in the chosen countries and calculate the similarity between the query suggestions' list at a period of time~$(t)$ with the query suggestions of the previous period~(\textit{t-1}) (see Equation \ref{jaccard_stability}).

\begin{equation}
\textit{Jaccard Stability} (A_{t},A_{t-1}) = \frac{|A_{t} \cap A_{t-1}|}{|A_{t} \cup A_{t-1}|} 
\label{jaccard_stability}
\end{equation}
 
\begin{figure}[t]
     \centering
     \includegraphics[width=8cm]{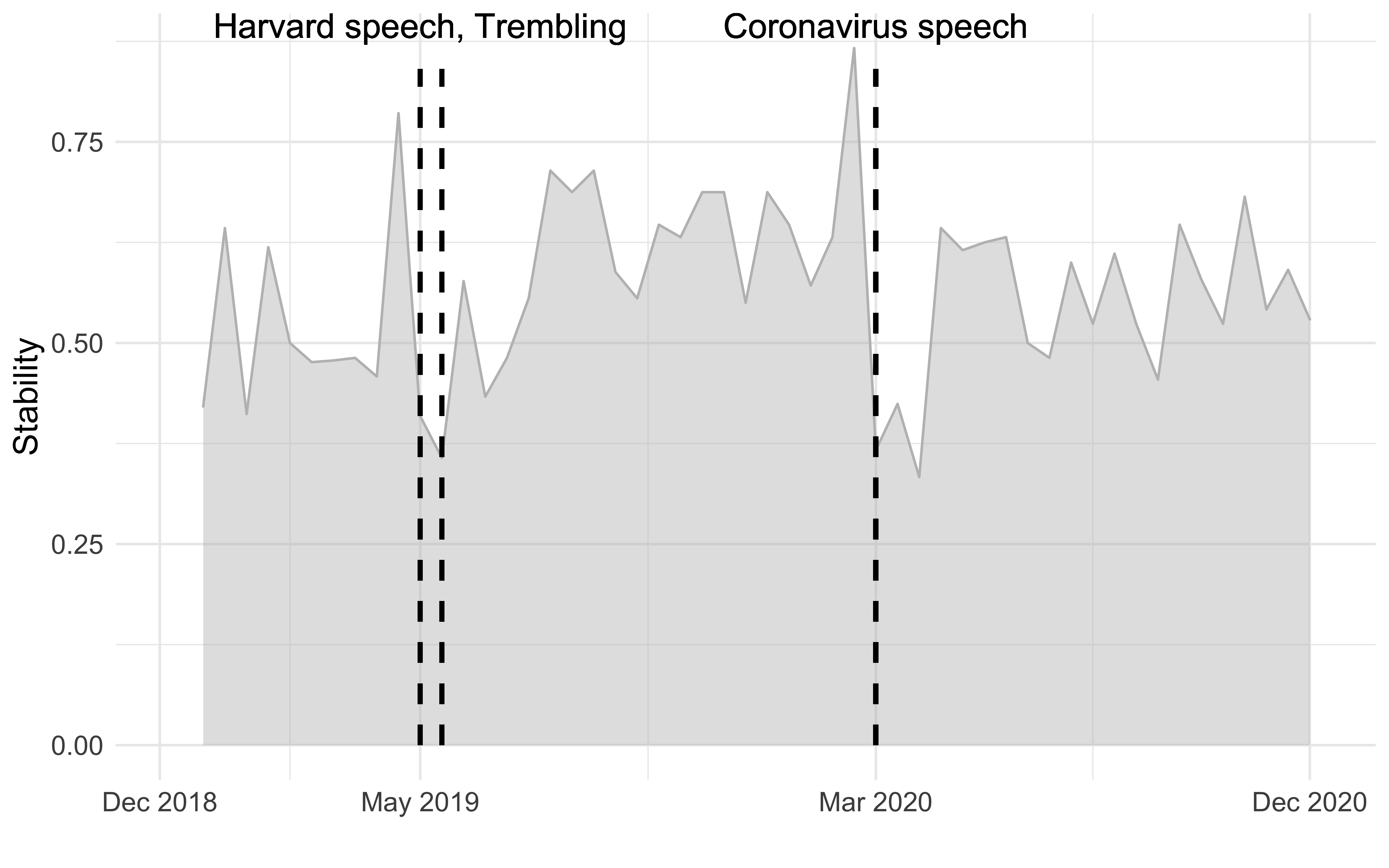}
     \caption{Stability of query suggestions (Jaccard Stability) for Angela Merkel over time in Germany.}
     \label{merkel_stability}
 \end{figure}

 Figure \ref{merkel_stability} shows the calculated stability of query suggestions for Angela Merkel in Germany, measured with the Jaccard Stability measure over time. The plot shows major drops in the stability score around June 2019 and March 2020, which coincide with the events mentioned previously. In June 2019, Angela Merkel suffered from trembling during a public speech
~\cite{bbc_germanys_2019} and was awarded an honorary doctoral degree at Havard university~\cite{die_bundesregierung_chancellors_2019, pazzanese_merkel_2019}. 
 The other notable change in March 2020 concurs with her nationwide TV speech on combating COVID-19 and taking the measures seriously~\cite{jones_germany_2020}. Subsequently, the query suggestions for Angela Merkel mirror interests in COVID-19 and related policies. 
 Looking at the case of Angela Merkel shows that the stability of query suggestions coincides with the public attention paid to her through major political events, and the measure serves as an indicator. 

\begin{figure}[t]
     \centering
     \includegraphics[width=8cm]{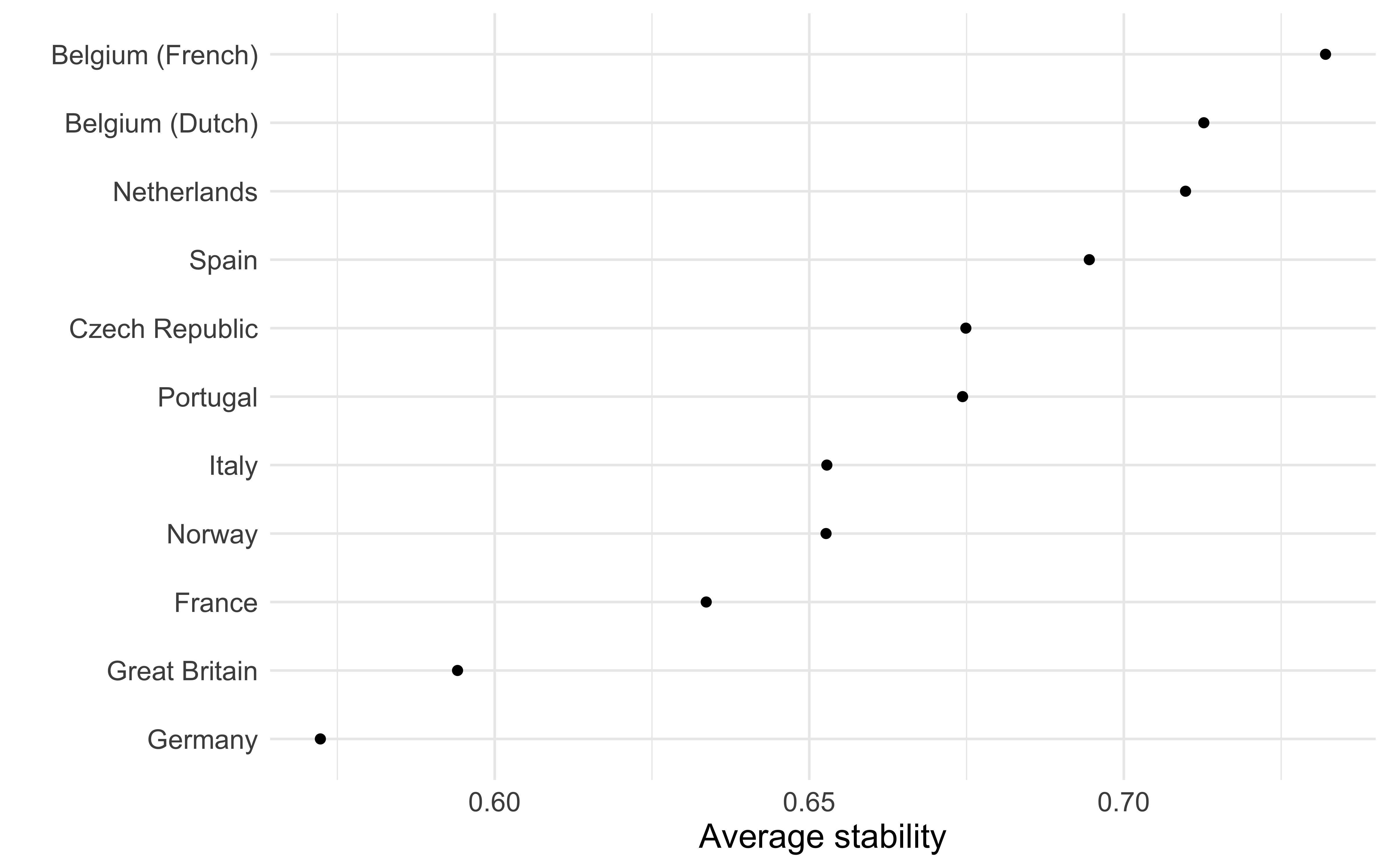}
     \caption{Average stability of query suggestions (Jaccard Stability) for Angela Merkel over all periods across countries.}
     \label{merkel_stability_countries}
 \end{figure}

As Figure \ref{merkel_stability_countries} shows, the average Jaccard stability of query suggestions for Angela Merkel over all periods varies strongly across countries. For example, the predictions for Angela Merkel are least stable in Germany, her country of origin, followed by Great Britain. The stability is lower (0.572 in Germany, 0.594 in Great Britain) than in other countries, like Belgium in French (0.732), the Netherlands (0.710), and Spain (0.695). Moreover, the Figure shows also shows that search queries in Belgium in French and Dutch are very similar to each other, even though the searches are conducted in different language settings.

\subsection{Measuring Similarity Across Countries}
Since the politicians' query suggestions approximately show what people have searched about them, we can explore how similar the interests are across countries and which factors can help to explain the similarities. 
We translated all query suggestions to English with Google Translate to analyze the similarity of query suggestions for politicians across countries.

To investigate the similarity in the search query suggestions between countries, we examine the Jaccard similarity coefficient of two lists of 2-weekly aggregated query suggestions of pairs of countries, A and B, for each period of time~$(t)$ (see Equation \ref{jaccard_countrycomparison}). In doing so, we follow the procedure of \cite{laufer_mining_2015}, who use the Jaccard similarity to analyze Wikipedia articles of the culinary domain to capture similarities of European food cultures.
We compare the query suggestions collected in the politicians' country of origin with those collected for the same politician in other countries.
We calculated the similarity of each two datasets of query suggestions by using R and the package purrr~\cite{henry_purrr_2019} to apply the Jaccard similarity measure. 

\begin{equation}
\textit{Jaccard Similarity} (A_{t}B_{t}) = \frac{|A_{t} \cap B_{t}|}{|A_{t} \cup B_{t}|} 
\label{jaccard_countrycomparison}
\end{equation}

\subsection{Meta-Attributes and Analytical Approach\label{sec:analytical_approach}}
As discussed in Section~\ref{sec:expectations}, we aim to investigate the correlation between a range of meta-attributes of politicians with the stability of their query suggestions over time and similarity across European countries.
We perform a multiple linear regression analysis to determine the relationships between the different meta-attributed groups and the stability as well as cross-country similarity. 
Moreover, we used the information on whether the search occurred in the politicians' country of origin as a dummy variable. 

We also investigate the political role of the politicians, i.e., whether the politician is part of a national cabinet, has the role of a party leader, or has been a lead candidate in the 2019 European Parliament election (c.f., Section~\ref{sec:Interest}). 
The attribute government party reflects the percentage of the observation period in which the politician is in the government party and not the opposition. For example, the value is 0.8 if a politician's party was in government for 80\% of the observation period. 
We measure the general politicization of European integration for each country by calculating the country-specific EU salience (see Equation~\ref{eu_salience}) based on data from the 2019~CHES expert survey on political parties~\cite{bakker_chapel_2019}. The measure accounts for the diverse party landscape, particularly in terms of the importance of the EU within countries, by weighting the average EU salience of a party $i$ with the party's vote share in the last national election~($votes_{i}$). 

  \begin{equation}
  EU Salience = \sum_{i=1}^{n} EU salience_{i} \cdot votes_{i} \\
  \label{eu_salience}
  \end{equation}

\section{Results}\label{sec:results}
This section presents the results of the multiple linear regression analyses performed to determine the relationships between the different meta-attribute groups and the stability (Section~\ref{sec:results_stability}) as well as cross-country similarity (Section~\ref{sec:results_variation}).

\subsection{Investigating the Stability of Query Suggestions}\label{sec:results_stability}

\begin{table}[t] \centering 
  \caption{Multiple linear regression of the average stability of politicians' query suggestions.} 
  \label{stability_results} 
  \small
\begin{tabular}{@{\extracolsep{5pt}}p{2cm}p{1cm}p{1cm}p{1cm}p{1cm}} 

\toprule
 & \multicolumn{4}{c}{\textit{Dependent variable:}} \\ 
\cline{2-5} 
\\[-1.8ex] & \multicolumn{4}{c}{Average stability of query suggestions over time} \\ 
\\[-1.8ex] & Model 1 & Model 2 & Model 3 & Model 4\\ 
\hline \\[-1.8ex] 
Search in country of origin & $-$0.139$^{***}$ & $-$0.139$^{***}$ & $-$0.139$^{***}$ & $-$0.139$^{***}$ \\ 
  & (0.005) & (0.005) & (0.005) & (0.005) \\ 
  & & & & \\
 \shortstack[l]{Party leader\\position} & $-$0.005$^{*}$ & $-$0.005$^{*}$ & $-$0.009$^{***}$ & $-$0.009$^{***}$ \\ 
  \shortstack[l]{(Reference:\\Cabinet)} & (0.002) & (0.002) & (0.002) & (0.002) \\ 
  & & & & \\ 
 Lead candidate & $-$0.070$^{***}$ & $-$0.070$^{***}$ & $-$0.071$^{***}$ & $-$0.071$^{***}$ \\ 
  \shortstack[l]{(Reference:\\Cabinet)}& (0.009) & (0.009) & (0.009) & (0.009) \\ 
  & & & & \\ 
  Government party & $-$0.016$^{***}$ & $-$0.016$^{***}$ & $-$0.017$^{***}$ & $-$0.017$^{***}$ \\ 
  & (0.002) & (0.002) & (0.003) & (0.003) \\ 
  & & & & \\
 Male & 0.010$^{***}$ & 0.010$^{***}$ & 0.012$^{***}$ & 0.013$^{***}$ \\ 
 \shortstack[l]{(Reference:\\Female)} & (0.002) & (0.002) & (0.002) & (0.002) \\ 
  & & & & \\ 
  EU salience &  & $-$0.001 & $-$0.001 & $-$0.001 \\ 
  &  & (0.0004) & (0.0004) & (0.0004) \\ 
  & & & & \\ 
  \shortstack[l]{Left-Right\\ideology} &  &  & $-$0.001 & $-$0.001 \\ 
  &  &  & (0.0005) & (0.001) \\ 
  & & & & \\ 
  \shortstack[l]{Left-Right\\ideology} &  &  &  & $-$0.0002 \\ 
  (squared) &  &  &  & (0.0002) \\ 
  & & & & \\ 
 Constant & 0.864$^{***}$ & 0.868$^{***}$ & 0.868$^{***}$ & 0.869$^{***}$ \\ 
  & (0.003) & (0.004) & (0.004) & (0.004) \\ 
  & & & & \\ 
\hline \\[-1.8ex] 
Observations & 6,721 & 6,721 & 6,402 & 6,402 \\ 
R$^{2}$ & 0.133 & 0.133 & 0.137 & 0.137 \\ 
Adjusted R$^{2}$ & 0.132 & 0.132 & 0.136 & 0.136 \\ 

\bottomrule
 \multicolumn{5}{p{1\columnwidth}}{\textit{Note:} Standard errors are shown in parentheses.  The Left-Right ideology is centered. The dependent variable is the average stability (Jaccard Stability) scores of politicians' search queries per country.   The sample is based on the politicians (N=793), their corresponding average stability scores (calculated based on 51 two-weekly intervals) from search queries per 10 distinct countries. Missing values were excluded from the regression if no query suggestions for politicians appeared in a search country or if they had missing values in the meta-attributes. $^{*}$p$<$0.05; $^{**}$p$<$0.01; $^{***}$p$<$0.001} 
\end{tabular} 
\end{table}

Table~\ref{stability_results} shows the results of four linear regression models investigating the stability of query suggestions for politicians within European countries.
Model 1 contains all meta-attributes, for which we have information for all of the politicians in the dataset. 
In models 2-4, we add meta-attributes according to our hypotheses (see Section \ref{sec:expectations}) into the regression analyses because the variables might be correlated with each other. 
We refrain from using time-fixed effects because we want to measure time differences, and we refrain from using politician-fixed effects because we want to measure the associations with politicians' meta-attributes. 

If the country of search matches the politician's country of origin, query suggestions are more volatile than in other countries. This result suggests that internet users search for more diverse information and receive more diverse query suggestions over time in the politicians' country of origin. 
Our results suggest that the role of politicians is associated with the stability of query suggestions over time. Query suggestions for politicians with a leading role are significantly less stable than those for politicians of national cabinets. At the same time, query suggestions for lead candidates of the 2019 European election are also significantly less stable and change more over time than those for politicians in the national cabinet and for party leaders. The search queries for these politicians might change more than those for other foreign politicians because these politicians were more visible during the campaign season. 
When the politicians belong to government parties, the query suggestions are significantly less stable than those of politicians in the opposition.
Query suggestions for male politicians are significantly more stable than those for female politicians.
The average EU salience of the search country is negatively associated with the stability of query suggestions, but the effect is not significant. 
Contrary to our expectations, we find no significant associations between the politicians' political ideology and the stability of query suggestions. This is also the case for the squared political ideology, which is used as a measure of political extremism. 

\paragraph{Further tests of latent political interest}
To further test whether query suggestions can be used as a latent measure of user attention paid to politicians, we also test the stability measure's construct validity. The examples for Angela Merkel (compare Section~\ref{sec:comp_anal}) show what we would expect: highly salient events result in new query suggestions immediately after the events. 
We systematically test whether query suggestions change more for elected lead candidates than those who were nominated but not elected for running as lead candidates in the European election. 
In the lead candidate process, European parties initially nominated politicians from their respective parties in a preliminary round to contend for the lead candidate position. Within their party, these nominated politicians were elected to represent the party and stand as candidates for the position of President of the European Commission (for more background information, see Section~\ref{sec:Interest}). 
As we know that lead candidates receive more media coverage and generally get more public attention once elected, we expect more volatile query suggestions for these politicians during the election period than for the unsuccessful lead candidate nominees.  We collected search query data for both groups, which allows us to further validate the stability of query suggestions as latent interest in politicians by testing whether the elected group with relatively more attention also has more volatile search queries.  

\begin{figure}
     \centering
     \includegraphics[width=8cm]{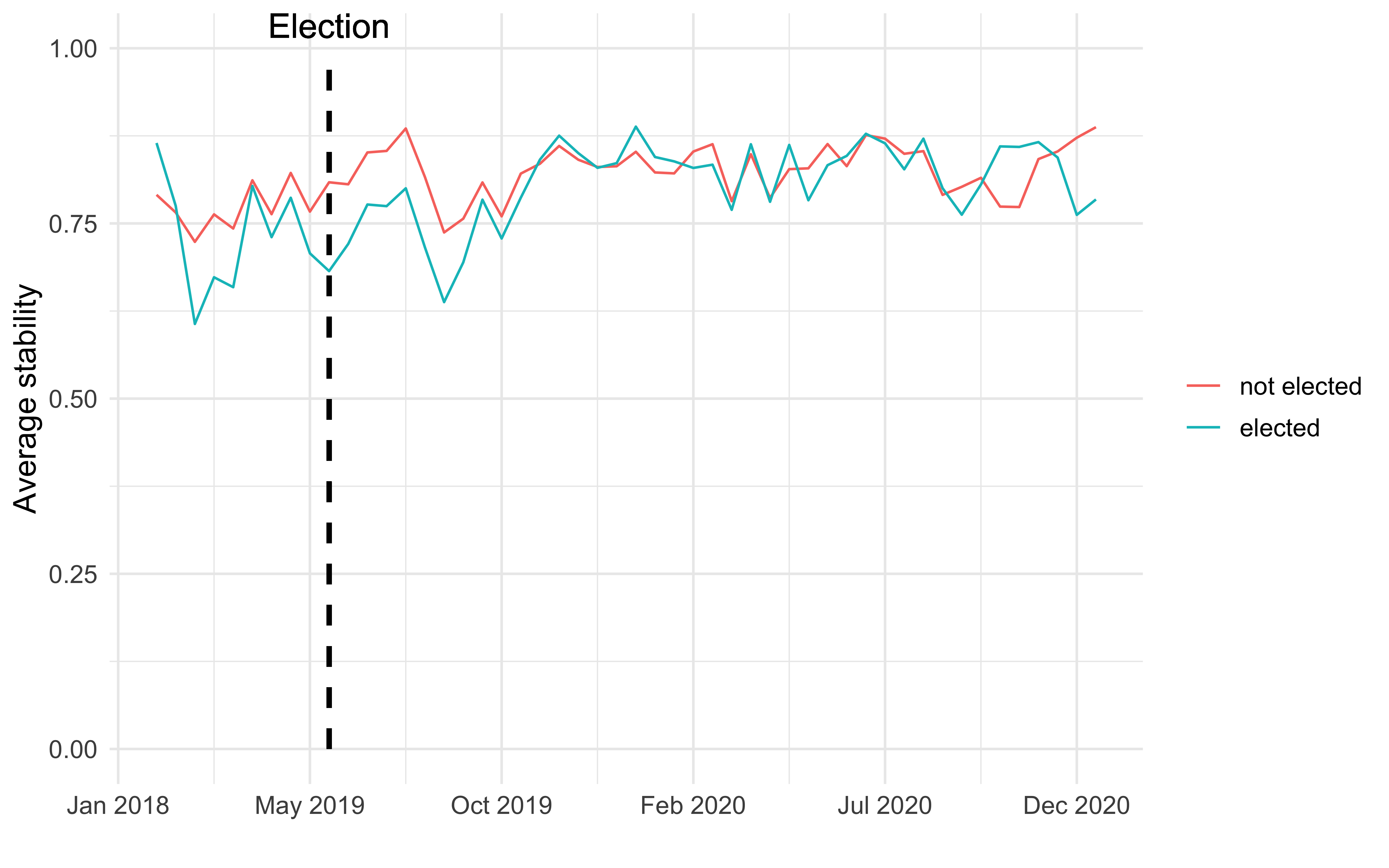}
     \caption{Average stability of query suggestions (Jaccard Stability) across countries for elected lead candidates (for running in the European election) and non-elected nominees for the lead candidate position.}
     \label{spitzenkandidaten_zeit}
 \end{figure}

Comparing the query suggestions of both groups in a linear regression model shows that the stability scores are significantly lower for the lead candidates than for the unsuccessful lead candidate nominees (see Table~\ref{spitzencand_test}). 
Looking at how the stability scores develop over time makes it apparent that the stability scores are notably lower for elected lead candidates around the European election 2019, while there are no observable differences otherwise (see Figure~\ref{spitzenkandidaten_zeit}). 

\begin{table}[t] \centering 
  \caption{Linear regression analysis results of the average stability of query suggestions for elected lead candidates (for running in the European election) and non-elected nominees for the lead candidate position.} 
  \label{spitzencand_test} 
\begin{tabular}{@{\extracolsep{5pt}}lc} 

\toprule
 & \multicolumn{1}{c}{\textit{Dependent variable:}} \\ 
\cline{2-2} 
\\[-1.8ex] & Stability of query suggestions \\ 
\hline \\[-1.8ex] 
 Elected lead candidates & $-$0.021$^{**}$ \\ 
 (Ref. Not elected) & (0.008) \\ 
  & \\ 
 Constant & 0.822$^{***}$ \\ 
  & (0.005) \\ 
  & \\ 
\hline \\[-1.8ex] 
Observations & 614 \\ 
R$^{2}$ & 0.011 \\ 
Adjusted R$^{2}$ & 0.009 \\ 
\bottomrule
 \multicolumn{2}{p{0.99\columnwidth}}{\textit{Note:} Standard errors are shown in parentheses.  
The dependent variable is the average stability (Jaccard Stability) scores per elected lead candidate and non-elected nominee for the lead candidate position. 
The sample is based on the politicians (N=14), their corresponding average stability scores (average across 10 countries) per 51 two-weekly intervals. Observations with missing values were excluded. The average stability scores for both groups are also visualized across time in Figure \ref{spitzenkandidaten_zeit}. 
 $^{*}$p$<$0.05; $^{**}$p$<$0.01; $^{***}$p$<$0.001}
\end{tabular} 
\end{table}

\subsection{Variation in Query Suggestions Across Countries}\label{sec:results_variation}

Table~\ref{similarity_results} displays the results of multiple linear regression analysis on the average similarity between the politicians' country of origin and other countries. 
The query suggestions have significantly greater similarity to those of other countries when they are about male politicians compared to when they are about female politicians. 
The more right the politicians are classified in terms of their political ideology, the significantly lower the similarity in query suggestions with other countries (see Model 2). 
In Model 3, squared political ideology was added. However, neither political ideology nor squared political ideology is significant. 
Contrary to our expectations, query suggestions for lead candidates in the 2019 European election show the least similarity with other countries compared to other political roles. 
We hypothesized that the query suggestions are more similar across countries for them than national party leaders and cabinet members due to their supranational role. However, the query suggestions were significantly more similar between countries for party leaders, followed by cabinet members, and least comparable for lead candidates (\textit{Spitzenkadidaten)}. 
When politicians were in the government during the observation period, query suggestions were significantly more similar across countries than when politicians were in the opposition. This suggests that interests in these politicians are generally more similar across Europe than for politicians in the opposition.

\begin{table}[t] \centering 
  \caption{Multiple linear regression results of the similarity of politicians' query suggestions between countries.} 
  \label{similarity_results} 
\begin{tabular}{@{\extracolsep{5pt}}llll} 
\toprule
 & \multicolumn{3}{c}{\textit{Dependent variable:}} \\ 
\cline{2-4} 
\\[-1.8ex] & \multicolumn{3}{c}{Average similarity of query suggestions} \\ 
\\[-1.8ex] & Model 1 & Model 2 & Model 3\\ 
\hline \\[-1.8ex] 
Party leader position  & 0.098$^{***}$ & 0.095$^{***}$ & 0.095$^{***}$ \\ 
  (Reference: Cabinet) & (0.011) & (0.012) & (0.012) \\ 
  & & & \\ 
 Lead candidate & $-$0.093$^{***}$ & $-$0.092$^{***}$ & $-$0.093$^{***}$ \\ 
  (Reference: Cabinet) & (0.026) & (0.026) & (0.026) \\ 
  & & & \\ 
 Male & 0.033$^{***}$ & 0.039$^{***}$ & 0.039$^{***}$ \\ 
   (Reference: Female) & (0.009) & (0.009) & (0.009) \\ 
  & & & \\ 
Government party & 0.049$^{***}$ & 0.062$^{***}$ & 0.054$^{***}$ \\ 
  & (0.012) & (0.012) & (0.014) \\ 
  & & & \\ 
Left-Right ideology &  & $-$0.011$^{***}$ & $-$0.010$^{***}$ \\ 
  &  & (0.002) & (0.002) \\ 
  & & & \\ 
Left-Right ideology &  &  & $-$0.001 \\ 
   (squared) &  &  & (0.001) \\ 
  & & & \\ 
 Constant & 0.257$^{***}$ & 0.247$^{***}$ & 0.257$^{***}$ \\ 
  & (0.014) & (0.014) & (0.017) \\ 
  & & & \\ 
\hline \\[-1.8ex] 
Observations & 2,650 & 2,600 & 2,600 \\ 
R$^{2}$ & 0.050 & 0.058 & 0.059 \\ 
Adjusted R$^{2}$ & 0.048 & 0.056 & 0.057 \\ 
\bottomrule
 \multicolumn{4}{p{0.99\columnwidth}}{\textit{Note:} Standard errors are shown in parentheses. 
 The Left-Right ideology is centered. The dependent variable is the average similarity (Jaccard Similarity) scores of politicians' search queries of their country of origin with the search queries of all other countries. 
 The sample is based on the politicians (N=267) who have a nationality of one of the  countries where search queries were collected and its corresponding average similarity with the 10 countries (calculated based on 51 two-weekly intervals). Observations with missing values were excluded. 
$^{*}$p$<$0.05; $^{**}$p$<$0.01; $^{***}$p$<$0.001}
\end{tabular} 
\end{table}

\section{Discussion}
 While we have recently witnessed an increase in the politicization in the European Union, less is known about cross-national aspects of online political information seeking related to EU issues and European politicians. In order to deepen our understanding of the dynamics within search engine query suggestions for European politicians, we examined as one part of this research the volatility of query suggestions for European politicians over time depending on important meta-attributes such as gender, political role, ideology, and government status.
The results presented in Section~\ref{sec:results} suggest that politicians' meta-attributes affect the diversity of search query suggestions over time, which leads to lower or higher stability scores.
Interestingly, the comparably higher volatility in query suggestions for lead candidates, influenced by search behavior, may suggest a growing politicization of the EU, as indicated in prior studies \cite{scherpereel_adoption_2017, hutter_politicizing_2019}, as users appear to seek on average more diverse information about them over time. However, this observation could also be attributed to a campaign effect. As these politicians were more visible during the EU election campaign, search queries may be more volatile during the campaign period but revert to a more stable pattern akin to other politicians post-campaign. 
Query suggestions are significantly more volatile when the search is performed in the country of origin of a politician. 
Nationality is a more relevant factor than even the political role of politicians.
This may indicate more (diverse) interest in national politicians and issues, even for politicians active in EU politics. Thus, the national identity plays a more significant role in determining users' attention to politicians. 
This finding is consistent with previous research that finds the politicization in the EU to be rather low in comparison to national politics~\cite{schneider_responsive_2018, rivas-de-roca_understanding_2021}.

Our research was also prompted by a second core question concerning the similarity of search query suggestions for European politicians across European countries and whether the similarity, or distinctiveness, varies depending on key meta-attributes of politicians like their gender, party role, government status, or ideology. 
Our analyses revealed that query suggestions for European politicians were most similar across countries if politicians held a leadership role, less similar if they were cabinet members, and least similar if they were lead candidates. 
Furthermore, they were less similar the more their party was classified on the right side of the political spectrum. The comparatively lower similarity in query suggestions for lead candidates and politicians with a right political ideology may be caused by heterogeneous news coverage of these politicians across countries that potentially caused different interests and search queries. This also aligns with a study indicating a heterogeneous presentation of lead candidates in Europe~\cite{schulze_spitzenkandidaten_2016}. Future research should tap into the similarity of news coverage across Europe and the link to online information searches. Moreover, while we have concentrated on similarity within and across countries, we have not characterized the qualitative changes in search engine predictions, which constitutes another subject for future investigation.

A limitation of our approach to analyzing the stability of suggested terms over time is that the Google search query suggestion generation mechanism is a black box system, which we cannot thoroughly probe. While we assume that the algorithms that provide the suggestion terms do not significantly change over two years, we cannot take that for granted. Google announced different so-called "Core Updates"\footnote{In an official tweet from Google's public liaison of search representative Danny Sullivan announced the major ``March 2019 Core Update'', also known as ``Florida 2'' was released to the public and might have an impact on the suggestion algorithm \cite{GoogleSearchLiaison2019}.} during the data acquisition phase. 
In Figure~\ref{merkel_stability} we can see a peak in stability in the weeks following this release, but a major update in the phase of the second drop did not produce any change. 
While we interpreted the changes in stability to be influenced by external events and not the underlying algorithms, which also seems reasonable when looking at the content, some uncertainty remains: since we only consider the first ten suggestions provided for each of the root queries, changes in less frequent query suggestions are not taken into account. 
Including more than the ten base suggestions could enable the detection of slighter changes.

In the analysis of the variation of query suggestions across countries, we use automated translations of query suggestions. This could be problematic in some cases, in which the translation is not correct or imprecise. Further, we do not consider the language in which a search is carried out and focus only on the content of the query. Investigating the correlation of the language of the query suggestion with regard to the country of origin and the country of search could deliver further insights.

In the scope of this paper, we have looked at the Google query suggestions as sets, disregarding their position. 
Considering the positions of query suggestions could yield further insights into the construct of interest in the search subject since changes in the order of query suggestions also suggest a change in the information need for the topic \citep{HaakS21}.
In this study, we focused on the Google search engine. Future research could shed light on the generalizability of the study findings to other platforms by collecting and analyzing search engine data on politicians across several search engine platforms. 
Moreover, it would be a fruitful direction for future research to look systematically into personalization of query suggestions based on search history and search location within countries. We technically avoided potential personalization effects in our research by performing searches always in a new automated browser environment. However, future research could thoroughly investigate the role of personalization in Google autocomplete across various country contexts and follow procedures of related studies on personalization in search engines \citep{hannak_measuring_2013, kliman2015location, robertson2018auditing}.

\section{Conclusion}
In this study, we investigate the effects of various meta-attributes on the stability over time and variation of search query suggestions for names of European politicians as indicators of latent political interest.
The results suggest that politicians' gender, political role, government party, and country of origin influence latent interest in politicians in online search, approximated by the Google search query suggestions. 
More specifically, we found that query suggestions were significantly more volatile over time when the search took place in the politicians' country of origin, when the politician fills an important political role, as in the case of lead candidates, and when politicians had a female gender identity. 
We found that being a governing party member, having a political leader role, and having a male gender identity were positively associated with a higher average similarity in query suggestions between countries. However, query suggestions were significantly less similar between countries when the politicians' party was classified more on the right of the political spectrum and when politicians were lead candidates than when they were in the national cabinet.

\begin{acks}
 This research was supported by the Digital Society research program funded by the Ministry of Culture and Science of the German State of North Rhine-Westphalia and the Deutsche Forschungsgemeinschaft (DFG, German Research Foundation) under Germany's Excellence Strategy – EXC 2126/1–390838866. We thank Heike Klüver and Simon Munzert for their valuable feedback on this work.
\end{acks}

\bibliographystyle{ACM-Reference-Format}
\bibliography{references}

\end{document}